\documentclass{article} 
\usepackage{spconf,amsmath,graphicx}
\usepackage{url}

\title{Detection of Audio-Video Synchronization Errors Via Event Detection}
\name{Joshua P. Ebenezer \sthanks{The first author worked on this during his internship at Amazon. His current affiliation is with the University of Texas at Austin.}, Yongjun Wu, Hai Wei, Sriram Sethuraman, Zongyi Liu }
\address{Amazon Prime Video, Seattle, WA}
\begin{document}
%
\maketitle
\begin{abstract}
We present a new method and a large-scale database to detect audio-video synchronization(A/V sync) errors in tennis videos. A deep network is trained to detect the visual signature of the tennis ball being hit by the racquet in the video stream. Another deep network is trained to detect the auditory signature of the same event in the audio stream. During evaluation, the audio stream is searched by the audio network for the audio event of the ball being hit. If the event is found in audio, the neighboring interval in video is searched for the corresponding visual signature. If the event is not found in the video stream but is found in the audio stream, A/V sync error is flagged. We developed a large-scaled database of 504,300 frames from 6 hours of videos of tennis events, simulated A/V sync errors, and found our method achieves high accuracy on the task. 
\end{abstract}
\begin{keywords}
Deep Learning, Database, Audio Video Synchronization
\end{keywords}
\section{Introduction}
Video quality is affected by a number of distortions that may occur at different stages of the video processing pipeline, from capturing to display. Most videos are accompanied by an audio stream that is synchronized with the visual stream. When the audio stream is not synchronized with the visual stream and is offset by a certain amount, humans are able to detect the offset and this negatively affect the user's quality of experience. These errors could occur during content capture, encoding, post-production, transmission, or play-back. Studies conducted by the International Telecommunication Union~\cite{itu} have found that humans typically find audio offsets of +50ms or more when audio is advanced with respect to (w.r.t.) the video stream, or -125 ms when audio is delayed w.r.t. vision, to be unacceptable. Zhao et al.~\cite{zhao} showed, through a human study, that humans find A/V sync errors to be even more detrimental to subjective experience than video impairments or audio impairments for virtual reality environments. Detecting such errors in A/V sync is a challenging task, but developing automated algorithms for it is vital for video-on-demand and livestreaming services to operate at scale. 
\par Previous efforts have focused on end-to-end training of the audio and the video against a sync/not-synced binary decision. Korbar et al.~\cite{korbar} proposed a two-stream network for the task. They defined ``easy" negatives as audio and visual segments that are from different videos, and ``hard" negatives as segments taken from the same video but offset by at least half a second. They reported poor performance on hard negatives and ``super-hard" negatives, which are segments offset by less than half a second. This is a major drawback in their approach. Khosravan et al.~\cite{khosravan} studied the use of attention models for the task and showed that spatio-temporal attention models can improve performance for end-to-end learning. The drawback in the method is that, by design, it cannot detect errors that are less than 2s in magnitude. Both training and evaluation sets have error magnitudes larger than 2s. Chung and Zissermann~\cite{chung} proposed an end-to-end network for lip-sync detection, but their method is not generalizable to A/V sync and is uniquely suited for lip-sync errors as it involves the tracking of the mouth in the video stream.
\par End-to-end paradigms suffer from the issues of not being able to predict small offsets and being difficult to interpret. Some other approaches~\cite{laurijssen,patent} propose the embedding of a unique signal into the audio and video streams, but this is not always feasible when one does not have control over content production. In this paper, we present a large-scale database of tennis videos and propose a novel method that is amenable to interpretation and can predict very small offsets with high accuracy. Videos of tennis events are labeled frame-by-frame as the events ``the ball is being hit by the racquet" (hit), ``the ball is bouncing" (bounce), and ``the ball is not in play or it is neither being hit nor is it bouncing" (neither). We create an audio event detector (AED) as a deep network that predicts whether segments in the audio stream correspond to a hit or not. We also create a video event detector (VED) as a deep network that predicts whether groups of frames in the visual stream correspond to the ball being hit or not. During testing, the AED searches the audio stream for a hit. If a hit is found, the neighbouring frames in the visual stream are queried by the VED on whether they contain a hit or not. If no hit is detected in the visual stream in the temporal neighbourhood of the detected hit in the audio stream, an A/V sync error is flagged. 

\section{Database creation}

Four videos of tennis events held in different courts were selected from the Amazon Prime Video catalog. Two videos were from WTA matches, and the other two videos were from ATP tours. The combined videos had a total duration of 6 hours and 504,300 frames.  We did not remove advertisements and kept shots of the crowd and replays as well, since we wanted to build a model that could generalize well. All videos were of 25 fps. We created a graphical user interface to label each frame as a ``hit" or a ``bounce" or ``neither". Frames of replays were labeled as ``neither" since the audio stream during replays generally contains commentary and not the audio of the play. Two labelers and an arbitrator labeled the frames.
\par The videos did not have A/V sync errors and hence the video labels were used to label the audio stream as well. The audio was captured at a sampling frequency of 48kHz in AAC-LC stereo. In order to account for the fact that the events we are trying to identify occur over a period of time, groups of three frames were considered as a single input and labeled with the label of the first of the three frames. At 25 fps, this corresponds to an interval between the first and third frames of 80ms. Audio segments of length 160ms were used as inputs to the AED, which at 48 kHz corresponds to 7680 pulse-code modulated (PCM) data points. Each segment was labeled by the label of the video frame that marked its beginning.  Out of the 504,300 frames collected in this way, only 2443 frames were hits. There was thus a data imbalance of approximately 1:200. 20\% of the data was set aside for testing. 80\% of the data was used for training and validation, out of which 80\% was used for training and the rest for validation.

\section{Methodology}

\begin{figure}
\includegraphics[width=\linewidth]{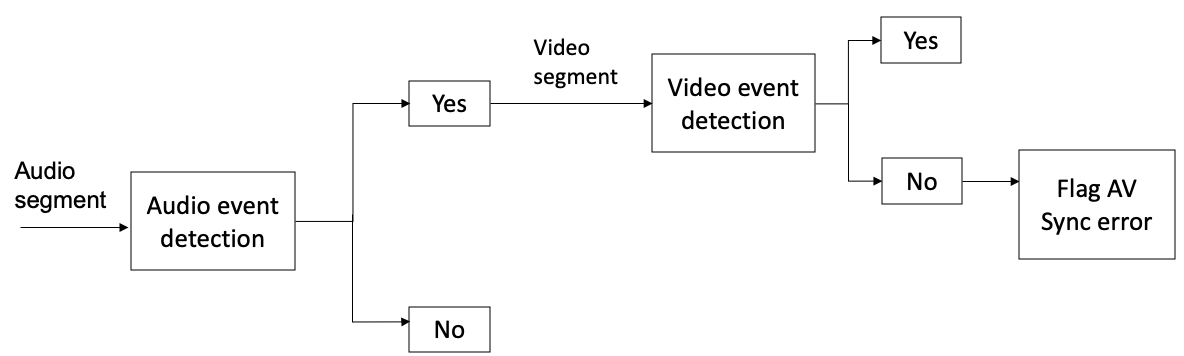}
\caption{Block diagram of the algorithm. The AED searches for the ball being hit in the audio stream. If it finds the event, the frames in the neighboring interval of time in the video stream are queried by the VED. If the event is not found in those frames, an A/V sync error is flagged.}
\label{fig:algo}
\end{figure}

The method that we propose to detect A/V sync errors in tennis videos is shown in Fig.~\ref{fig:algo}. In the following section, we describe the architecture and design of the audio event detector and the video event detector.
\subsection{Audio Event Detector}

\subsubsection{Audio processing}

Each audio segment is of length 7680 audio data points with a sampling frequency of 48 kHz. The mel-frequency cepstral coefficients are extracted from each segment. 2048 audio data points are used for each window, and windows are spaced by 128 audio data points. 61 MFCCs were computed for each window, so that the data formed a $61\times60$ input frame. The first and second temporal derivatives of the MFCCs were also computed. The input to the AED was thus a image of dimensions $61\times 60\times 3$.

\subsubsection{Training}

A ResNet~\cite{resnet} pre-trained on the ImageNet~\cite{imagenet} database was used as the audio event detector. Training CNNs on MFCCs or log-mel-spectrograms has been shown to produce state-of-the-art results in audio processing for a number of applications~\cite{bird1,bird2,speech1,audiodeep}. Pretraining CNNs on images has been shown to transfer well to audio tasks and has been used to establish benchmarks in a number of audio datasets~\cite{audioimageproof,audioimagefirst,audioimagesecond,audioimagethird,audioimagefourth}. Segments of audio were fed to the network and trained against a binary decision of ``hit/not a hit". The binary cross entropy loss was used. Early stopping was implemented by stopping the training process when the precision on the validation data dropped for three consecutive epochs. High precision is desirable for the audio event detector because if a false positive ``hit" is detected by the network when there is actually no hit present, and the VED searches for a hit and does not find it, a false AV sync flag is raised. In our application scenario, operators have to manually check the stream if a sync error is raised, and frequent false flags can cause fatigue and irritation. Besides this, hits occur at a high frequency during rallies, and missing a few hits is therefore tolerable as long as a sufficient number of hits are detected within the interval of play. The network is trained on the entire training set. The input segments are separated by 40 ms, which is the time interval between two frames in the visual stream and is thus the smallest interval at which events can be detected. The audio segment is of length 160 ms because we found that the sound of some hits sometimes extends to that amount of time. However, because of the large amount of overlap between adjacent windows, we also found that a number of false positives were adjacent (in time) to false negatives during evaluation. We discuss how we resolved this in section III.

\subsection{Video Event Detector}
\subsubsection{Video Preprocessing}

\begin{figure}
\includegraphics[width=\linewidth]{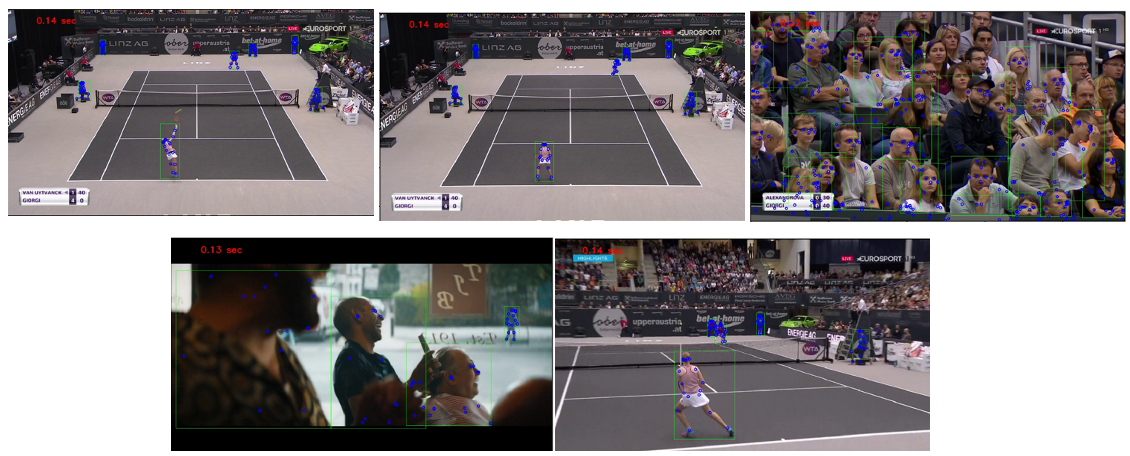}
\caption{Representative frames from the database after bounding box and pose detection are applied.}
\label{fig:pose}
\end{figure}

The video frames were passed through the pose detector proposed in~\cite{hrnet}. The pose detector marks the position of joints and also provides a bounding box for persons in the frame. The scenes that are found in the broadcast video are diverse and include shots of the crowd, advertisements, replays, etc. It is therefore impossible to predict the number of persons that can be in a frame, and so the frames with the bounding boxes are passed as-is to the subsequent network. Scenes from a video with the bounding boxes and pose overlaid are shown in Fig.\ref{fig:pose}. The frames were then resized to size $960\times 540$ to keep computational costs reasonable and normalized before passing them to the network. Data augmentation was performed by randomly flipping the frames horizontally. Three frames are sent to the network at a time, and all three would have the label of the first frame in that group.

\subsubsection{Training}

We use a mixed-convolution network~\cite{mc} for the video event detector. The first 6 layers of the network are from the C3D~\cite{c3d} network pre-trained on UCF101~\cite{ucf}. At the end of 6 layers, the temporal dimension is compressed completely and subsequent layers are 2D convolutional residual blocks. Three 2D convolutional residual blocks of 512 layers each are used after the initial 3D convolutional layers taken from C3D. 
\par We found that the video network could not learn to identify tennis hits from the video if it was trained on the whole dataset. Training typically did not converge well because of the massive data imbalance. This was probably because the visual signature of the ball being hit occupies a small part of the frame (spatially) and the rest of the frame appears similar to many other frames in the video where the ball is in play but not being hit. The features and patterns during ball hits in video frames are not as prominent as those in audio data, where a clear difference in features exists between ``hits" and ``not hits". For example, when players are running in the tennis court whether there is a ball hit or not, those (many) frames look very similar to each other, whereas in the audio stream the sound of the ball being hit is distinctly heard over the background noise of play. The other difficulty is that we are targeting identification within a very small window in time (3 frames or 80 ms). Most video activity recognition algorithms test over much larger temporal durations, generally 25 frames or above. The data imbalance also exacerbates the problem since the network can classify all frames as "not a hit" and achieve a low and stagnant loss. Random undersampling, random oversampling, and loss reweighing did not work to solve the issue since in this case the evaluation set has the same data imbalance as the training set.

\begin{figure}
\centering
\includegraphics[height=3cm]{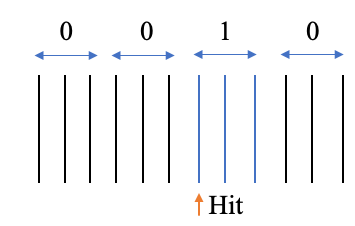}
\caption{Illustration of labeling and sample selection for the third stage of VED training. Each vertical line represents a video frame. Positive samples are blocks of 3 frames each starting from frames labeled as ``hits". Negative samples are blocks of 3 frames each that are immediately adjacent to positive samples from 6 frames behind to 6 frames ahead of the hit}
\label{fig:third}
\end{figure}

We first tried training the VED against the \textit{entire} database and using the VED in parallel with the AED, in order for the VED to detect hits separately from the AED and then correlate the hit detector outputs of the VED and AED across time. This was not feasible due to the aforementioned difficulties. Instead, we found that the AED could be used as a reliable filtering mechanism, and that training the VED to distinguish frames near the hit was sufficient. We trained the VED in three stages, following ~\cite{korbar}. In the first stage, we train with randomly chosen negative (i.e. ``not hits") examples from the training set and with \textit{all} the positive examples in the training set, such that the data is balanced. In the second stage, we train all the positives with harder negatives, which we define as blocks of frames corresponding to when the tennis ball is bouncing. Such frames always show the ball in play and the players in motion, which force the network to learn the features necessary to distinguish a hit from a scene when the ball is in play.  In the third and final stage, we train with blocks of frames that are immediately adjacent to the positives. An example of negatives for the third stage of training is shown in Fig.~\ref{fig:third}.

The VED was trained with an early stopping mechanism. Training was stopped after recall on the validation set fell for three consecutive epochs. It is important to maintain high recall for the VED because if it fails to identify hits where there are actually hits, the final AV sync detector will raise a number of false flags, which is undesirable.

\subsection{A/V synchronization error detector}

\begin{figure}
\centering
\includegraphics[height=3cm]{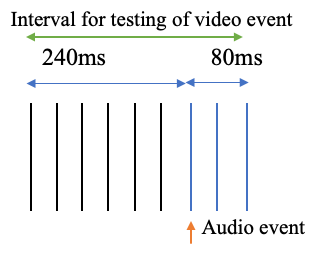}
\caption{Illustration of search space for the VED. If the AED detects a hit, the VED searches the frames in the video in the 6 frames preceding the time instance at which the audio event was detected and in the three frames afterwards.}
\label{fig:search}
\end{figure}

We test blocks of three frames each, which at 25fps corresponds to 40 ms. We therefore search for hits in the video stream between 240 ms before the hit is detected in the audio, and 80 ms after. An illustration is shown in Fig.~\ref{fig:search}. There are three non-overlapping groups of three frames each in this interval, and the VED searches each of these separately to find a hit. There are thus three predictions made by the VED for each audio segment identified as a hit by the AED, and if any of those predictions are a ``hit", no AV sync error is flagged.

\section{Experiments and results}

\subsection{Hyperparameters}

Both the AED and the VED were trained with the Adam optimizer with decay rates $\beta_1=0.9$ and $\beta_2=0.99$. An exponentially decaying learning rate scheduler was used with a decay coefficient of 0.95 for both networks. 

\subsection{Audio Event Detector}

The confusion matrix for the AED on the test set is shown in Table~\ref{tab:aed1}. We found that most false positives were immediately adjacent to false negatives. Out of the 145 false positives, 108 audio segments were immediately adjacent (i.e. within $\pm 40$ ms) to false negatives. Most of these errors are caused by the fact that the auditory signature of hits extend over multiple segments. These errors are automatically corrected by the overall system for AV sync error detection because the VED searches for hits in the \textit{neighborhood} of the time at which the hit was detected in the audio. Table~\ref{tab:aed2} shows the confusion matrix for the AED, adjusted with adjacent false positive and false negative errors being treated as true positives or true negatives. With the adjusted result, the effective precision of the AED is 90.2\% and the recall is 72.76\%, which fully satisfies the requirements of our application scenario.

\begin{table}[htbp]
\caption{Confusion matrix for AED}
\begin{center}
\begin{tabular}{|c|c|c|c|}
\hline
& Positives & Negatives \\
\hline
 Predicted Positives & 234 & 145  \\ 
 \hline
 Predicted Negatives & 236 & 100217 \\ 
 \hline
\end{tabular}
\label{tab:aed1}
\end{center}
\end{table}

\begin{table}[htbp]
\caption{Adjusted Confusion matrix for AED}
\begin{center}
\begin{tabular}{|c|c|c|c|}
\hline
& Positives & Negatives \\
\hline
 Predicted Positives & 342 & 37  \\ 
 \hline
 Predicted Negatives & 128 & 100325 \\ 
 \hline
\end{tabular}
\label{tab:aed2}
\end{center}
\end{table}

\subsection{AV Synchronization Error Simulation}

The VED only evaluates the segments that are declared as hits by the AED. In order to simulate AV sync errors, we randomly introduced offsets between the video and audio for half of the samples that were predicted as positives by the AED. The offset was a random number of video frames selected from between $-15$ and $+15$, excluding the range from $-3$ to $+6$. The range from $-3$ frames to $+6$ frames was skipped because the frames in this interval would be searched for the video event.

\subsection{AV Synchronization Error Detection}

Table~\ref{tab:avsync} shows the confusion matrix for the final task of AV sync error detection. The segments chosen by the AED (with errors simulated as described in the earlier subsection) were sent to the VED, which inspected the neighboring frames and made predictions accordingly. 100,832 audio segments were to be evaluated in the test set. There were 470 hits among these. Out of these, 379 were detected as hits by the AED. 342 of these are actually hits (after adjustment). Out of the 379, 155 examples were assigned an AV sync error by the random process described earlier. All 379 examples were queried by the VED. If no hit was found in the neighborhood of the chosen audio segment, an AV sync error was flagged. Precision obtained by the VED is 81.25\% and the recall is 83.87\%.

\begin{table}[htbp]
\caption{Confusion matrix for AV Sync Error}
\begin{center}
\begin{tabular}{|c|c|c|c|}
\hline
& Positives & Negatives \\
\hline
 Predicted Positives & 130 & 30  \\ 
 \hline
 Predicted Negatives & 25 & 138 \\ 
 \hline
\end{tabular}
\label{tab:avsync}
\end{center}
\end{table}

\section{Conclusion}

We proposed a novel method and created a large-scale database for detecting errors in A/V synchronization of tennis videos. This method is able to detect very small errors that are just outside the human threshold of perception and establishes a new state-of-the-art in a relatively unexplored area. We intend to explore alternative processing techniques for the audio and video streams to further advance the state-of-the-art. The excellent results that we have achieved give impetus for extending this idea to other sports and for high-frame-rate contents.

{\small
\bibliographystyle{IEEE}
\bibliography{conference_041818}
}
\end{document}